\documentclass[11pt, a4paper]{amsart}
\usepackage{amsfonts}

\usepackage[dvips]{epsfig}
\usepackage{amsmath}
\usepackage{amsthm}
\usepackage{amsbsy}
\usepackage{amsgen}
\usepackage{amscd}
\usepackage{amsopn}
\usepackage{amstext}
\usepackage{amsxtra}
\usepackage{dsfont}

\newcommand {\T} {\mathcal{T}}
\newcommand {\M} {\mathcal{M}}
\newcommand {\Z} {\mathbb{Z}}
\newcommand {\R} {\mathbb{R}}

\newcommand {\E} {\mathbb{E}}

\newcommand{\var}{\operatorname{Var}}
\newcommand {\sphere}{\mathcal{S}^2}

\newcommand {\eigspc} {\mathcal{E}_n}
\newcommand {\eigspcdim} {\mathcal{N}}
\newcommand {\eigval} {E_{n}}
\newcommand{\length}{\mathcal{Z}}

\newtheorem{theorem}{Theorem}[section]

\newtheorem{lemma}[theorem]{Lemma}

\newtheorem{conjecture}[theorem]{Conjecture}

\newtheorem{principle}[theorem]{Principle}

\begin{document}
\title{On the nodal lines of random and deterministic Laplace eigenfunctions}
\author{Igor Wigman}
\address{Department of Mathematics, Cardiff University, Wales, UK}
\email{wigmani@cardiff.ac.uk}
\maketitle

\begin{abstract}

In the present survey we present some of the recent results concerning the geometry
of nodal lines of random Gaussian eigenfunctions (in case of spectral degeneracies)
or wavepackets and related issues. The most fundamental example, where the spectral
degeneracy allows us to consider random eigenfunctions (i.e. endow the eigenspace with
Gaussian probability measure), is the sphere, and the corresponding eigenspaces are
the spaces of spherical harmonics; this model is the primary focus of the present survey.
The list of results presented is, by no means, complete.

\end{abstract}

\section{Introduction}

Nodal patterns (first described by Ernest Chladni in 18th century) appear in
many problems in engineering, physics and the natural sciences: they describe the
sets that remain stationary during vibrations, hence their importance in such
diverse areas as musical instruments industry, mechanical structures, earthquake study
and other areas. They also arise in the study of wave propagation, and in astrophysics;
this is a very active and rapidly developing research area.

\subsection{Some basic notation}

Let $(\M,g)$ be a compact Riemannian surface (for example $S^{2}$, the two
dimensional unit sphere equipped with the round metric),
and $\Delta$ be the Laplace-Beltrami operator on $\M$. We are interested in the eigenvalues $\lambda$
and the corresponding eigenfunctions $\phi$ of $-\Delta$, so that $$\Delta \phi + \lambda\phi = 0.$$
In case $\M$ has a boundary, we impose either the Dirichlet boundary condition $\phi|_{\partial M} \equiv 0$,
or the Neumann boundary condition $$\frac{\partial \phi}{\partial \nu}|_{\partial \M} \equiv 0,$$ or any mixture
of the conditions above.
The general spectral theory states that there is a complete orthonormal basis of $L^2(\M)$ which
consists of eigenfunctions, i.e. we may choose a sequence of functions
$$\{\phi_{j}:\M\rightarrow\R \}_{j=1}^{\infty}$$ and
corresponding nondecreasing sequence of eigenvalues $\{ \lambda_{j} \}_{j=1}^{\infty}$ so that the orthonormal set
$\{\phi_{j}\}$ spans the whole of $L^{2}(\M)$. Note that we allow multiple eigenvalues i.e. spectral
degeneracies.

Let $\phi:\M\rightarrow\R$ be any real valued function. The nodal line of $\phi$ is its zero set $$\phi ^{-1}(0) = \{x\in \M:\: \phi(x)=0 \}.$$ In principle, a nodal line might have self-intersections\footnotemark; however generically
it is a smooth $1$-dimensional curve with components homeomorphic to either the circle (``closed component")
or an interval (in this case it must intersect the boundary; it is called an ``open component").
We are interested in the geometry of the nodal lines of $\phi_{j}$ as $j\rightarrow\infty$. The most basic aspect of the nodal line is, of course, its length; let us denote $l_{j}$
to be the length of the nodal line of $\phi_{j}$.
In this survey we will only consider the $2$-dimensional case; however most of the results
presented extend for higher dimensions.

\footnotetext{For example the eigenfunction
$\phi(x,y) = \sin(10\pi x) \sin(20\pi y),$ defined on the torus $T^{2} = \R^{2}/\Z^{2}$, having eigenvalue
$\lambda=500\pi^2$.}

\subsection{Yau's conjecture and Berry's RWM}
\label{sec:Yau and Berry}

Yau conjectured ~\cite{Y1,Y2} that for any smooth $\M$, $l_{j}$ are commensurable to $\sqrt{\lambda_{j}}$ for any {\em smooth} metric $g$ in the sense that there exist two constants $c(\M),C(\M)$ so that
\begin{equation}
\label{eq:Yau}
c(\M) \sqrt{\lambda_{j}} \le l_{j} \le C(\M) \sqrt{\lambda_{j}}
\end{equation}
for every $j\ge 1$.
The lower bound was proved by Bruning and Gromes
~\cite{Bruning-Gromes} and Bruning ~\cite{Bruning} for the planar
case. Donnelly and Fefferman
~\cite{Donnelly-Fefferman} finally settled Yau's conjecture for real
analytic metrics. However, in its full generality, Yau's conjecture is still open.

In his seminal work ~\cite{Berry 1977}, Berry argued that the high energy behaviour of the eigenfunctions
should be universal, at least for ``generic" chaotic surfaces $\M$. He proposed to compare an eigenfunction
with eigenvalue $\lambda$ to a ``typical" instance of an isotropic, monochromatic random wave
with wavenumber $$k=\sqrt{\lambda}$$ (nowadays called Berry's Random Wave Model - RWM). A $1$-dimensional
version of the random wave was used by Rice in order to investigate the likelihood of a given signal
to exceed a level. Longuet-Higgins generalized Rice's model to $2$-dimensional plane
to describe the movement of the sea and ocean waves.

There are several ways to construct the ensemble of random waves associated with wavenumber $k$. One way to do it
is consider summations of type\footnotemark
\begin{equation*}
u_{k;J}(x) = \frac{1}{\sqrt{J}}\Re \left( \sum\limits_{j=1}^{J} e^{k\langle\theta_{j}, x \rangle + \phi_{j}}\right),
\end{equation*}
on $\R^{2}$, where $\theta_{j}$ are random directions drawn uniformly on the unit circle, and $\phi_{j} \in [0,2\pi)$
are the random phases. One would like to define the random wave $u_{k}(x)$ on $\R^{2}$ as the limiting ensemble
\begin{equation*}
u_{k}(x) = \lim\limits_{J\rightarrow\infty} u_{k;J}(x);
\end{equation*}
which should converge in distribution. Another mathematically rigorous way to define the
wavenumber $k$ Random Wave is to identify it as the unique Gaussian
isotropic random field (ensemble of functions) with covariance function
\begin{equation}
\label{eq:cov RWM}
r_{RWM}(x,y) = J_{0}(k|x-y|),
\end{equation}
where $J_{0}$ is the usual Bessel function.

\footnotetext{In reality the summation is slightly more complicated than the one presented - see e.g.
~\cite{Bogomolny Percolation}, (1).}

Since, according to the RWM, the random waves model the high-energy eigenvalues, the nodal lines of
random waves should also model the nodal lines of honest eigenfunctions. This approach allows us to study
local quantities like the nodal length, boundary intersections and intersections with a test curve etc.

Suppose, for example, we are interested in the nodal length on the torus. Then we are to choose a representative
planar domain $U\subseteq\R^{2}$ (e.g. a rectangle with the same aspect ratio and area as the torus), and study
the distribution of $\length_{U;\sqrt{\lambda}}$, the nodal length of random wave with wavenumber
$k=\sqrt{\lambda}$ inside $U$.
It is easy to compute the expected length to be of order of magnitude
$$\E[\length_{U;\sqrt{\lambda}}] \sim const \cdot \sqrt{\lambda} |U|,$$ where $|U|$ is the area
of $|U|$, and Berry argued ~\cite{Berry 2002} that the variance should be of order
\begin{equation}
\label{eq:nod len var RWM}
\var\left(\length_{U;\sqrt{\lambda}}\right) \sim const \cdot |U|\log{\lambda}.
\end{equation}

\subsection{Bogomolny and Schmit's percolation model}

\label{sec:Bogomolny}

The RWM, however, does not help if one is interested in making predictions regarding the more subtle
(and arguably, more interesting) aspects of the nodal structures such as
the number of nodal domains\footnotemark, their size distribution, the size of the largest nodal domain,
the inner radius etc. For this purpose an elegant independent bond percolation-like model was introduced by Bogomolny
and Schmit ~\cite{Bogomolny Percolation}.
According to this model, the nodal domains should correspond to the clusters connected by open bonds, and the nodal line corresponds to these clusters' boundaries ~\cite{Bogomolny SLE}. Let $\nu_{j} = N(\phi_{j})$ be the number of nodal domains of $\phi_{j}$, and $N(u_{\sqrt{\lambda}};U)$ the number of nodal domains of random wave $u_{\sqrt{\lambda};U}$ on $U$.
The main criticism against this model is that the independence assumption ignores all the dependencies that occur between the bonds; to try to justify the independence assumption, Bogomolny and Schmit ~\cite{Bogomolny Harris} apply a heuristic principle, the so-called Harris criterion.

\footnotetext{The nodal domains are the connected components of the complement of the nodal line.}

By the classical Courant Nodal Domain Theorem (see e.g. ~\cite{Courant Hilbert}), $\nu_{j} \le j$, and Pleijel ~\cite{Pleijel} asymptotically improved the latter to
\begin{equation*}
\limsup\limits_{j\rightarrow\infty} \frac{\nu_{j}}{j} \le 0.691\ldots.
\end{equation*}
On the other hand, no nontrivial lower bound for $\nu_{j}$ could be found, since one may find a sequence
of energy levels $\lambda_{j_{k}}$ on the torus (say), so that the corresponding eigenfunctions would have
only $2$ nodal domains.
Bogomolny and Schmit ~\cite{Bogomolny Percolation} used the general percolation theory to predict that $$N(u_{\sqrt{\lambda}};U)$$ should be asymptotically Gaussian, with mean and variance proportional to $$|U| \cdot \left(\sqrt{\lambda}\right)^2 = |U|\cdot \lambda.$$ More strikingly, in their later paper, Bogomolny et al. ~\cite{Bogomolny SLE} argue that since, according to the recent developments in the percolation theory (see e.g. Smirnov ~\cite{Smirnov}), the ``interface" (cluster boundaries)
should converge to $SLE_{6}$, the distribution of the largest\footnotemark
component of the nodal line (rather than its length) should converge to $SLE_{6}$.

\footnotetext{According to the percolation theory, there exists exactly one component that ``covers" the
whole domain; this is the only macroscopic component of the nodal line. }

\subsection{Equidistribution conjecture}

It is conjectured\footnote{In a recent survey by S. Nonnenmacher ~\cite{NN},
this conjecture was attributed to S. Zelditch.}
that on any chaotic surface $\M$ (for example, any negatively curved surface,
or any ergodic billiard), the nodal lines are asymptotically equidistributed in $\M$, so that, in particular, the nodal length has asymptotic shape of type
\begin{equation}
\label{eq:lj sim c*sqrt(lam)}
l_{j} \sim c_{\M} \cdot \sqrt{\lambda_{j}}
\end{equation}
(refinement of Yau \eqref{eq:Yau}), for some constant $c_{\M} > 0$.
Despite the fact that heuristically, \eqref{eq:lj sim c*sqrt(lam)} follows from the RWM
(it follows for example from the results mentioned in the end of Section \ref{sec:Yau and Berry}), this conjecture seems extremely difficult or even out of reach by the present analytic methods,
and it seems highly unlikely that it is going to be settled in the near future.

However, some information could be inferred from the completely integrable case,
even though the picture that emerges here is very different.
For instance, one may use the spectrum degeneracy
of the standard torus to
easily construct sequences of eigenfunctions $\phi_{n_{1;j}}$ and $\phi_{n_{2;j}}$,
$i=1,2$, $j=1,2,\ldots$, so that
\begin{equation*}
\mathrm{length} \left( \phi_{n_{i;j}} ^{-1}(0)\right) \sim c_{i}\cdot \sqrt{\lambda_{n_{j}}},
\end{equation*}
$i=1,2$, with $c_{1}\ne c_{2}$; one may obtain such sequences on the sphere only slightly modifying
the same argument.
One way to infer some information is use the following heuristic principle:
\begin{principle}[``Word exchangeability"]
Any property satisfied by {\em generic} eigenfunctions on (all) completely integrable manifolds is
also satisfied by {\em all} eigenfunctions on a {\em generic} chaotic manifold.
\end{principle}

\subsection{Acknowledgements}

The author would like to thank Ze\'ev Rudnick for suggesting to consider some of the problems in this survey,
and having many deep and fruitful conversations. I would also like to thank Mikhail Sodin for many extremely
stimulating and fruitful discussions regarding some of the subjects that appear in this survey and their context,
and many useful comments on an earlier version of this manuscript.
In addition, it is important for me to acknowledge the organizers of the Dartmouth International Conference
in Spectral Geometry for organizing such a wonderful conference, that included many wonderful talks and speakers,
in addition to the free and relaxed informal environment or atmosphere, stimulating and encouraging collaboration and new ideas' exchange, and also for the generous financial support.


\section{Some results}

\subsection{Spherical harmonics}

It is well known that the eigenvalues $E$ of the Laplace
equation
\begin{equation*}
\Delta f +E f = 0
\end{equation*}
on the $2$-dimensional sphere $\sphere$ are all the numbers of the form
\begin{equation}
\label{eq:eigval def} \eigval = n(n+1),
\end{equation}
where $n$ is an integer. The corresponding
eigenspace is the space $\eigspc$ of spherical harmonics of
degree $n$; its dimension is $$\eigspcdim_{n} = 2n+1.$$
Given an integer $n$, we fix an $L^2(\sphere)$ orthonormal
basis of $\eigspc$
\begin{equation*}
\eta_{1}^{n} (x), \, \eta_{2}^{n}
(x),\ldots ,\eta_{2n+1}^{n} (x),
\end{equation*}
giving an identification $\eigspc\cong\R^{\eigspcdim_{n}}$. For further reading on the spherical harmonics we refer the reader to ~\cite{AAR}, chapter 9.

\subsection{Random models}

In case of spectral degeneracy, such as the sphere or the torus, we may consider
a {\em random eigenfunction} lying inside an eigenspace. For the sphere, we define it as
\begin{equation}
\label{eq:rand eigfnc def} f_{n}(x)= \sqrt{\frac{2}{\eigspcdim_{n}}}
\sum\limits_{k=1}^{\eigspcdim_{n}} a_{k}\eta^{n}_{k}(x),
\end{equation}
where $a_k$ are standard Gaussian $N(0,1)$ i.i.d. That is, we use the
identification $$\eigspc\cong\R^{\eigspcdim_{n}}$$ to endow the space $\eigspc$ with
Gaussian probability measure $\upsilon$ as
\begin{equation*}
d\upsilon(f_{n}) = e^{-\frac{1}{2}\|\vec{a}\|^2}\frac{da_{1} \cdot\ldots\cdot
da_{\eigspcdim_{n}}}{(2\pi)^{\eigspcdim_{n}/2}},
\end{equation*}
where $\vec{a} = (a_{i})\in\R^{\eigspcdim_{n}}$ are as in \eqref{eq:rand eigfnc def}.

Note that $\upsilon$ is invariant with respect to the
orthonormal basis for $\eigspc$. Moreover, the Gaussian {\em random field} $f^{m}_{n}$ is
isotropic in the sense that for every $x_{1},\ldots x_{l}\in\sphere$ and every orthogonal
$R\in O(3)$,
\begin{equation}
\label{eq:f(Rx1...Rxl)d=f(x1,...xl)}
\left(f_{n}(Rx_{1}),\ldots, f_{n}(Rx_{l})\right) \stackrel{d}{ =} \left(f_{n}(x_{1}),\ldots, f_{n}(x_{l})\right).
\end{equation}
There exists yet another way to define $f_{n}$: it is the centered Gaussian isotropic random field with
covariance function
\begin{equation}
\label{eq:covar spher Pn}
r_{n}(x,y) := \E [f_{n}(x) \cdot f_{n}(y)] = P_{n}(\cos d(x,y)),
\end{equation}
where $P_{n}$ are the well-known Legendre polynomial of degree $n$, and $d$ is the spherical distance.
The Legendre polynomials admit Hilb's asymptotics
\begin{equation}
\label{eq:Hilb}
P_{n}(\cos(\phi)) \approx \sqrt{\frac{\phi}{\sin\phi}}J_{0}(\phi(n+1/2)),
\end{equation}
i.e. almost identical to RWM \eqref{eq:cov RWM}, up to the  ``correction factor" $\sqrt{\frac{\phi}{\sin\phi}}$.
This factor seems to ``keep a trace" or ``remember" about the geometry of the sphere.

For generic manifolds there are no spectral degeneracies, so that we should consider linear Gaussian combinations
of individual eigenfunctions (``wavepackets").
The two most accepted models are the so-called {\em long energy window} and
{\em short energy window}. In the long window case we consider Gaussian combinations of eigenfunctions with
eigenvalue lying in the window $[0,\lambda]$ for $\lambda\rightarrow\infty$
\begin{equation*}
f_{\lambda}^{L}(x) = \frac{1}{\sqrt{N_{\M}(\lambda)}}\sum\limits_{\sqrt{\lambda_{j}}\le \lambda} a_{j} \phi_{j}(x),
\end{equation*}
where $x\in\M$, and $$N_{\M}(\lambda) = \#\{\lambda_{j}\le \lambda\}$$ is the spectral function;
the reason we took the square root of the eigenvalues is that it makes it more convenient to write the short energy
window random function
\begin{equation*}
f_{\lambda}^{S}(x) = \frac{1}{\sqrt{N_{\M}((\sqrt{\lambda}+1)^2)-N_{\M}(\lambda)}}\sum\limits_{\sqrt{\lambda} \le \sqrt{\lambda_{j}}\le \sqrt{\lambda}+1} a_{j} \phi_{j}(x).
\end{equation*}
The short energy model\footnote{The window $[\lambda,\lambda+1]$ may be replaced by $[\lambda,\lambda+a]$ for
any constant $a>0$.} is considered more significant, as it is more representative of the individual eigenfunctions \footnotemark; however working with the long energy window is relatively easier. The spectral prefactor in the definition of $f^{L,S}$ was introduced to make the expected $L^2$-norm unity.

\footnotetext{Much like the random trigonometric polynomials $\sum\limits_{n=N}^{N+\sqrt{N}}a_{n}\cos(nt)$ on $[0,2\pi]$ is more representative of $\cos(Nt)$ than $\sum\limits_{n=1}^{N}a_{n}\cos(nt)$, where in both summations $a_{n}$ are standard Gaussian i.i.d.; for example it possesses asymptotically the same number of zeros.}

As usual, for any random variable
$X$, we denote its expectation $\E X$. For example, with
the normalization factor in \eqref{eq:rand eigfnc def}, for every $n$
{\em fixed} point $x\in\sphere$, one has
\begin{equation}
\label{eq:E(f(x)^2)=1} \E [f_{n}(x)^2] =
\frac{\sphere}{\eigspcdim_{n}}\sum\limits_{k=1}^{\eigspcdim_{n}} \eta^{n}_k(x) ^2 = 1,
\end{equation}
a simple corollary from the Addition Theorem (see
~\cite{AAR}).

Any characteristic $X(L)$ of the nodal line
$$f_{n}^{-1}(0)=\{x\in\sphere:\: f_{n}(x)=0 \} $$ is a random variable. The most natural
characteristic of the nodal line of $f_{n}$ is, of course, its
length $\length(f_{n})$.
One may then study the distribution of the random variable $\length(f_{n})$ for a random Gaussian
$f_{n}\in\eigspc$, as $n\rightarrow\infty$. It is also natural to consider the number $N(f_{n})$ of
the nodal domains of $f_{n}$, i.e. the connected components of $$\sphere\setminus f_{n}^{-1}(0);$$
its distribution should be consistent to the one predicted by Bogomolny and Schmit based on their percolation model
(see Section \ref{sec:Bogomolny} above).

\subsection{Some generic results}

Berard ~\cite{Berard}, and subsequently Zelditch ~\cite{Z} found that the expected nodal length of the long energy window random functions is
\begin{equation*}
\E\left[ \length(f^{L}_{\lambda}) \right] \sim const\cdot \sqrt{\lambda},
\end{equation*}
consistent with Yau. Zelditch ~\cite{Z} also extended this result to the short energy window case
\begin{equation*}
\E\left[ \length(f^{S}_{\lambda}) \right] \sim const\cdot \sqrt{\lambda}.
\end{equation*}
In addition, for $g$ real analytic, Zelditch ~\cite{Z} considered the complexified manifold
$(\M_{\mathbb{C}}, g_{\mathbb{C}})$ (whose projection on $\Im {z} = 0$ is $(M,g)$), the
analytic continuations $\phi_{j}^{\mathbb{C}}$, and the corresponding random combinations
$f^{\mathbb{C};L,S}_{\lambda}$, defined analogously to the real random combinations $f^{L,S}_{\lambda}$.
In this case the zeros are isolated points in $\M_{\mathbb{C}}$; Zelditch ~\cite{Z} found that
their expected number is again proportional to $\sqrt{\lambda}$.

Toth and Wigman ~\cite{TW1} considered the number of boundary intersections $$\mathcal{I}\left(f^{L,S}_{\lambda}\right)$$
of the nodal line of $f^{L,S}_{\lambda}$
in case $\M$ is a generic billiard (i.e. a planar oval with a smooth boundary), or equivalently (up to the factor $2$), the number of open components. They found the correct order
of magnitude for the expected number of intersections to be
\begin{equation*}
\E\left[\mathcal{I}\left(f^{L,S}_{\lambda}\right)\right] \sim const |\partial \M| \sqrt{\lambda},
\end{equation*}
where $|\partial \M|$ is the boundary length of $\M$, and the constants differ in the long and short window cases.
This result is consistent to both Yau's conjecture, random wave model, and the resulting interpretation of
the boundary trace as approximating trigonometric polynomials; the asymptotics depends only on the boundary length of
the billiard, notably independent of its shape.

\subsection{Number of nodal domains}


Nazarov-Sodin ~\cite{Nazarov Sodin} found the correct order of magnitude for the expected number of nodal domains
of random spherical harmonics, and established an exponential decay result for deviations from the mean.

\begin{theorem}[Nazarov-Sodin ~\cite{Nazarov Sodin}]
\label{thm:Nazarov-Sodin}
There exists a constant $a>0$ so that the expected number of nodal domains is asymptotic to
\begin{equation}
\label{eq:nod dom expected NS}
\E[N(f_{n})] = an^2 + o(n^2).
\end{equation}
Moreover, for every $\epsilon > 0$, there exist two constants $c(\epsilon),C(\epsilon) > 0$, so that
\begin{equation}
\label{eq:nod dom exponential dec}
\Pr\left( \left|\frac{N(f_{n})}{n^2}-a \right| > \epsilon \right) \le C(\epsilon) e^{-c(\epsilon)n}.
\end{equation}
\end{theorem}

The result \eqref{eq:nod dom expected NS} on the expected number of nodal domains is of more general nature:
it extends to a wide range of sequences of random fields (to appear in a paper by Nazarov and Sodin).
For example, rather than taking a random element lying in a single spherical harmonics space,
one may superpose elements from several spaces). However, unlike the rapid decay \eqref{eq:nod dom exponential dec}
in the particular case of spherical harmonics, in the more general situation, Nazarov-Sodin's result
does not prescribe the rate of decay of the tails of the distribution. Instead, they prove the weaker statement:
for any $\epsilon > 0$
\begin{equation*}
\lim\limits_{n\rightarrow\infty} \Pr\left( \left|\frac{N(f_{n})}{n^2}-a \right| > \epsilon \right) = 0.
\end{equation*}

One disadvantage of the results above and the method of their proofs is the fact that the constant
$a>0$, whose existence is established, remains mysterious and completely open;
one cannot establish the dependence of $a$ depends on the underlying random field.
As an example, one may oppose $f_{n}$ to
$g_{n}$, a superposition of random spherical harmonics of degree $\le n$. A generalized
version of Nazarov-Sodin's Theorem implies the existence of a constant so that
$$\E[N(g_{n})] \sim bn^2.$$ Theorem \ref{thm:Nazarov-Sodin} and its generalization do not shed a light on the
the relation between $a$ and $b$.

Though the expectation result is consistent with the percolation model, Theorem \ref{thm:Nazarov-Sodin}
gives us no clue what would be the result for the variance. However, it provides us with a very
strong rate of decay (namely exponential); the decay results are usually complementary to the variance results.
The authors also proved that the prescribed rate of decay cannot be improved, so that the exponential decay
they establish, is of the correct order of magnitude.

\subsection{Nodal length of random spherical harmonics}

It is a standard application of the Kac-Rice formula (see e.g. ~\cite{CL}) to compute the expected nodal length of random spherical harmonics ~\cite{Berard}
\begin{equation}
\label{eq:Elen=*sqrt(E)}
\E \left[\length(f_{n}) \right] = c\cdot \sqrt{\eigval},
\end{equation}
where
\begin{equation*}
c = \sqrt{2}\pi,
\end{equation*}
(see also ~\cite{Neuheisel} and ~\cite{Wig1}). Our main concern in this pursue is the subtle
question of the variance.

Based on the natural scaling of the sphere (e.g. the relation to the Legendre polynomials,
in particular \eqref{eq:covar spher Pn}), we conjectured ~\cite{Wig1}, that
\begin{equation*}
\var(\length(f_{n})) \sim const\cdot n.
\end{equation*}
Surprisingly, the variance turned out to be much smaller, due to an unexpected cancellation (``Berry's cancellation phenomenon"). We derived the following asymptotics for the nodal length variance, improving the earlier
bounds of Neuheisel ~\cite{Neuheisel} and Wigman ~\cite{Wig1}:
\begin{theorem}[Wigman ~\cite{Wig2}]
\label{thm:var length} As $n\rightarrow\infty$, one has
\begin{equation}
\label{eq:var length} \var\left(\length(f_{n})\right) =
\frac{65}{32}\log{n}+O(1).
\end{equation}
\end{theorem}

Note that the leading constant $\frac{65}{32}$ in \eqref{eq:var length} is different\footnotemark from the one
predicted by Berry for the RWM (see \eqref{eq:nod len var RWM}). Our explanation for this discrepancy
is the nontrivial local geometry of the sphere. It seems reasonable that for a generic
chaotic surface, the nodal length variance for $f^{L,S}_{\lambda}$ should be logarithmic; the leading
constant is then an artifact of the local geometry. One of our central goals is to find this dependency
explicitly, namely, given a Riemannian surface, $(\M,g)$, compute the leading constant in front
the logarithm (if this prediction is indeed correct).

\footnotetext{Since $f_{n}$ is odd for odd $n$ and even for even $n$, the nodal lines are invariant w.r.t.
the involution $x\mapsto -x$. Therefore the natural planar domain to compare would be one of area of
a hemisphere rather than of the full sphere. }

Theorem \ref{thm:var length} implies that the series
\begin{equation*}
\sum\limits_{n=1}^{\infty} \var\left(  \frac{\length(f_{n})}{\E[\length(f_{n})]}  \right)
\end{equation*}
of variances of the normalized length $$\frac{\length(f_{n})}{\E[\length(f_{n})]},$$ is convergent.
Together with the Borel-Cantelli Lemma, it implies that for independently chosen $f_{n}$,
\begin{equation*}
\lim\limits_{n\rightarrow\infty}\frac{\length(f_{n})}{\sqrt{\eigval}} = c,
\end{equation*}
almost surely, where $c>0$ is the same constant as in \eqref{eq:Elen=*sqrt(E)}.

The same problem may be also considered on the standard $2$-dimensional torus $\T = \R^{2}/\Z^{2}$.
Here the eigenvalues are of the form $$ \eigval^{\T} = 4\pi^2 n ,$$ where
$n$ is an integer, expressible as a sum of two integer squares, and the corresponding eigenspace
is spanned\footnotemark by functions $\cos\left(2\pi\langle \lambda,\, x   \rangle   \right)$ and
$\sin\left(2\pi\langle \lambda,\, x   \rangle   \right)$, where $\lambda\in\Z^{2}$ with
$\| \lambda\|^2=n$, are all the lattice points lying on the circle of radius $\sqrt{n}$;
its dimension is $r_{2}(n)$, the number of representations of $n$ as a sum of two squares.
The Gaussian random eigenfunction is a stationary random field with the covariance
function
\begin{equation}
\label{eq:rnT}
r_{n}^{\T}(x) = \frac{1}{\eigspcdim}\sum\limits_{\|\lambda \|^2 = n}
\cos\left( 2\pi \langle \lambda,\, x \rangle  \right).
\end{equation}

\footnotetext{Note the invariance w.r.t. $\lambda\mapsto -\lambda$, so that we need to factor
the set of lattice points by $\pm$.}


It is again standard to compute that the expected nodal length of this ensemble to be proportional
to $\sqrt{\eigval^{\T}}$ ~\cite{RW}, and we are interested in the asymptotic behaviour of
the variance again. This question was initially considered by
Rudnick and Wigman ~\cite{RW}; however it got only a partial answer then. An (almost) complete answer
will be given in the forthcoming paper Krishnapur-Kurlberg-Wigman ~\cite{KKW}.

Even though we have
an explicit expression \eqref{eq:rnT} for the covariance function, no analogue of \eqref{eq:Hilb} is
known for the asymptotic long-range behaviour of $r_{n}^{\T}$ (recall \eqref{eq:covar spher Pn}).
As a replacement, we cope with some subtle issues of the arithmetics of lattice points lying
on a circle. Here as well we observed the ``arithmetic Berry's cancellation", a phenomenon of
different appearance but similar nature to ``Berry's cancellation phenomenon".

\subsection{Level exceeding}

Let us define the \emph{spherical harmonics}
\emph{level exceeding} measure as follows: for all $z\in (-\infty ,\infty ),$
\begin{equation}
\Phi _{n}(z):=\int_{S^{2}}\mathds{1}(f_{n}(x)\leq z)dx\text{,}
\label{empmea}
\end{equation}%
where $\mathds{1}(\cdot )$ is, as usual, the indicator function which takes
value one if the condition in the argument is satisfied, zero otherwise. In
words, the function $\Phi _{n}(z)$ provides the (random) measure of the set
where the eigenfunction lie below the value $z.$ For example, the value of
$\Phi _{n}(z)$ at $z=0$ is related to the so-called defect
\begin{equation*}
\mathcal{D}_{n}:=\mathrm{meas}\left( f_{n}^{-1}(0,\infty )\right) -\mathrm{%
meas}\left( f_{n}^{-1}(-\infty ,0)\right)
\end{equation*}%
by the straightforward transformation
\begin{equation*}
\mathcal{D}_{n}=4\pi -2\Phi _{n}(0).
\end{equation*}%
Of course, $4\pi -\Phi_{n}(z)$ provides the area of the excursion set $$
\mathcal{A}_{n}(z):=\left\{ x:f_{n}(x)>z\right\} .$$
Clearly, for all $z\in \mathbb{R}$,
\begin{equation*}
\mathbb{E}\left[ \Phi _{n}(z)\right] =4\pi \Phi (z),
\end{equation*}%
where $\Phi (\cdot )$ is the cumulative distribution function of the standard
Gaussian. The following lemma (see Marinucci-Wigman ~\cite{MW1}) deals with the variance of
$\Phi_{n}(z)$ as $n\rightarrow \infty $.

\begin{lemma}
\label{lem:var(Phi_l)} For every $z\in \mathbb{R}$,
\begin{equation*}
\var(\Phi_{n}(z)) = z^2 \phi(z)^2 \cdot \frac{1}{n} + O_{z}\left(
\frac{\log{n}}{n^2} \right),
\end{equation*}
where $\phi$ is the standard Gaussian probability density function.
\end{lemma}

In particular, for $z\neq 0$, Lemma \ref{lem:var(Phi_l)} gives the
asymptotic form of the variance as $n\rightarrow \infty $. In contrast, for $
z=0$ (this case corresponds to the defect), this yields only a $``o"$-bound
and one needs to work harder to obtain a precise estimate; we do so in the forthcoming
paper:
\begin{theorem}[Marinucci-Wigman ~\cite{MW2}]
As $n\rightarrow\infty$, the defect variance is asymptotic to
\begin{equation*}
\var(\mathcal{D}_{n}) \sim C \cdot \frac{1}{n^2},
\end{equation*}
where $C>0$ is some constant.
\end{theorem}

In light of Lemma \ref{lem:var(Phi_l)} it is then natural to normalize $\Phi_{n}(z)$
and define the \emph{spherical harmonics empirical process} by
\begin{equation}
G_{n}(z):=\sqrt{n}\left[ \int_{S^{2}}\mathds{1}\left( f_{n}(x)\leq z\right)
dx-\left\{ 4\pi \times \Phi (z)\right\} \right]  \label{emppro}
\end{equation}
for $n=1,2,...$, $z\in (-\infty ,\infty)$.

In ~\cite{MW1} we proved the following result:
\begin{theorem}[Marinucci-Wigman ~\cite{MW1}]
\label{thm:Gl->Ginf} (The Uniform Central Limit Theorem) As $n\rightarrow
\infty ,$ the process $G_{n}(z)$ converges in distribution to $G_{\infty}(z)$,
where $G_{\infty}(z)$ is the mean zero, degenerate Gaussian process on $\R$
given by $$G_{\infty }(z)=z\mathbb{\phi }(z)Z$$ with
$Z\sim N(0,1)$ standard Gaussian random variable.
\end{theorem}

This result, in particular, implies the full asymptotic dependence of $G_{n}(z)$ for different values of $z$,
as $n\rightarrow\infty$.
See the next Section for some explanation to this phenomenon.

\subsection{Nodal line vs. Level curves}

Interestingly, the behaviour of level curves $f_{n}^{-1}(L)$ for $L\ne 0$ is very different compared to
the behaviour of nodal lines. Let $\length^{L}(f_{n})$ be the length of the level curve $f_{n}^{-1}(L)$.
It is standard to compute the expected length, using the Kac-Rice formula
\begin{equation*}
\E[\length^{L}(f_{n})] = c_{1}e^{-L^2/2}\sqrt{E_{n}}
\end{equation*}
consistent with the nodal case $L=0$. However, unlike the nodal lines, level length variance
is asymptotic to \cite{Wig3}
\begin{equation}
\label{eq:lenL(fn)=L4n}
\var(\length^{L}(f_{n})) \sim c_{2}L^4 e^{-L^2}\cdot n;
\end{equation}
it is also interesting to observe the fact that the leading term depends on $L^4$ (a priori, the
dependence on $L$ should be symmetric w.r.t. $L\mapsto -L$, however we would rather expect $L^2$;
its dependence cancels out - another obscure cancellation related to this problem).

Moreover, the length of the level curves is asymptotically fully correlated, in the sense that\footnotemark{}
\begin{equation}
\label{eq:rho(L1,L2)=1-o(1)}
\rho(\length^{L_{1}}(f_{n}), \length^{L_{2}}(f_{n})) = 1-o_{n\rightarrow\infty}\left(  1 \right).
\end{equation}
Let us relate between the latter and the setup of Theorem \ref{thm:Gl->Ginf}.
One may express $\Phi_{n}(z)$ in terms of
the level lengths using
\begin{equation}
\label{eq:Phi(z)=int(levels)}
\Phi_{n} (z) = \int\limits_{-\infty}^{z} \length^{L}(f_{n}) dL.
\end{equation}
Intuitively, the asymptotic degeneracy of $\Phi_{n}(z)$ for different values of $z$ is then
an artifact of the asymptotic full dependence \eqref{eq:rho(L1,L2)=1-o(1)} of the individual
values in the integrand on the RHS of \eqref{eq:Phi(z)=int(levels)}.

\footnotetext{For two random variables $X$, $Y$ the correlation is defined as $\rho(X,Y) =
\frac{Cov(X,Y)}{\sqrt{\var(X)} \sqrt{\var(Y)}}$; $|\rho|\le 1$ measures the linear correlation
between $X$ and $Y$.}

One possible explanation for the phenomenon \eqref{eq:rho(L1,L2)=1-o(1)} is the
following conjecture, due to Mikhail Sodin (see Marinucci-Wigman
~\cite{MW2} for further reading on this conjecture). For $x\in \sphere$ and $L\in\mathbb{R}$
let $$\length^{L}_{x}=\length^{L}_{x}(f_{n})$$ (the ``local length")
be the (random) length of the unique component\footnotemark of $f_{n}^{-1}(L)$
that contains $x$ inside (or $0$, if $f_{n}$ does not cross the level $L$).

\footnotetext{We should assume that $f_{n}(x) \ne L$; the latter is satisfied almost surely.}

\begin{conjecture}[M. Sodin]
\label{conj:loc len full dep} The local lengths are asymptotically fully
dependent in the sense that for every $x\in \sphere$ and $L_{1},L_{2} \in\mathbb{R}$
\begin{equation*}
\rho\left(\length^{L_{1}}_{x}, \length^{L_{2}}_{x}\right) =
1-o_{n\rightarrow\infty}(1).
\end{equation*}
\end{conjecture}

Intuitively, it should be clear that Conjecture \ref{conj:loc len full dep}
implies \eqref{eq:rho(L1,L2)=1-o(1)} (and thus also the asymptotic degeneracy of the
level exceeding measure via \eqref{eq:Phi(z)=int(levels)}), since $\length^{L}$
is some summation of $\length^{L}_{x}$ over some $x$ on the sphere.

All the results above piece up nicely together while dealing with the corresponding questions
in case of {\em Gaussian subordinated} random fields\footnote{The author would like to thank Domenico Marinucci for discussing this subject.}: let $G:\R\rightarrow\R$
be a (possibly nonlinear) nice function and define the random field $g_{n}$ by $$g_{n}(x) = G(f_{n}(x));$$
$g_{n}$ is Gaussian subordinated. We are interested in the nodal length of $g_{n}$
\begin{equation*}
\length_{g_{n}} = \mathrm{length}(g_{n}^{-1}(0)),
\end{equation*}
so that if $z_{1},\ldots, z_{k}$ are all the zeros\footnotemark of $G$. It is obvious that
\begin{equation*}
\length_{g_{n}} = \sum\limits_{i=1}^{k} \length^{z_{i}}_{n}.
\end{equation*}

Therefore, the expected nodal length of $g_{n}$ is
\begin{equation*}
\E[\length_{g_{n}}] = c_{1} \sum\limits_{i=1}^{k} e^{-L^2/2}\sqrt{E_{n}}
\end{equation*}
for some explicitly given $c_{1} >0$.

\footnotetext{We do allow infinitely many zeros of $G$; it is easy to modify the formulas to follow
in this case.}

It is not difficult to see that
\eqref{eq:lenL(fn)=L4n} together with \eqref{eq:rho(L1,L2)=1-o(1)} gives an elegant and compact asymptotic result
(for $n\rightarrow\infty$) for the nodal length variance of $g_{n}$ as
\begin{equation*}
\var[\length_{g_{n}}] \sim c_{2} \left( \sum\limits_{i=1}^{k} e^{-z_{i}^2/2} z_{i}^2  \right)^2 \cdot n,
\end{equation*}
with some explicit $c_{2} > 0$, provided that $z_{i}\ne 0$ for at least one index $i$.


\begin{thebibliography}{99}

\bibitem{AAR}
Andrews, George E.; Askey, Richard; Roy, Ranjan {\em Special functions
Encyclopedia of Mathematics and its Applications} 71. Cambridge
University Press, Cambridge, 1999.

\bibitem{Berard}
B\'erard, P. {\em Volume des ensembles nodaux des fonctions propres
du laplacien}.   Bony-Sjostrand-Meyer seminar, 1984--1985, Exp. No.
14 , 10 pp., \'Ecole Polytech., Palaiseau, 1985.

\bibitem{Berry 1977}
Berry, M. V. {\em Regular and irregular semiclassical
wavefunctions}. J. Phys. A  {\bf 10}  (1977), no. 12,
2083--2091.


\bibitem{Berry 2002}
Berry, Michael V.  {\em Statistics of nodal lines and points in chaotic
quantum billiards: perimeter corrections, fluctuations, curvature}
J. Phys. A {\bf 35} (2002), 3025--3038.

\bibitem{Bogomolny Percolation}
Bogomolny, E; Schmit, C.;
Percolation model for nodal domains of chaotic wave functions, Phys. Rev. Lett. 88, 114102 (2002).

\bibitem{Bogomolny Harris}
Bogomolny, E.; Schmit, C. Random wavefunctions and percolation. J. Phys. A 40 (2007), no. 47, 14033–-14043.

\bibitem{Bogomolny SLE}
Bogomolny, E.; Dubertrand, R.; Schmit, C. SLE description of the nodal lines of random wavefunctions. J. Phys. A 40 (2007), no. 3, 381–-395.

\bibitem{Bruning}
J. Br\"{u}ning {\em \"{U}ber Knoten Eigenfunktionen des
Laplace-Beltrami Operators}, Math. Z. 158 (1978), 15--21.


\bibitem{Bruning-Gromes}
J. Br\"{u}ning and D. Gromes {\em \"{U}ber die L\"{a}nge der
Knotenlinien schwingender Membranen}, Math. Z. 124 (1972), 79--82.


\bibitem{Courant Hilbert}
Courant, R.; Hilbert, D. Methods of mathematical physics. Vol. I. Interscience Publishers, Inc., New York, N.Y., 1953. xv+561 pp.

\bibitem{CL}
Cram\'{e}r, Harald; Leadbetter, M. R. Stationary and related stochastic processes.
Sample function properties and their applications. Reprint of the 1967 original.
Dover Publications, Inc., Mineola, NY, 2004.

\bibitem{Donnelly-Fefferman}
H.~Donnelly, and C.~Fefferman; Nodal sets of eigenfunctions on
Riemannian manifolds, Invent. Math. {\bf 93} (1988), 161--183.



\bibitem{KKW}
Krishnapur, M., Kurlberg P., Wigman I.;
Fluctuations of the nodal length of random eigenfunctions of the Laplacian on the arithmetic torus,
in preparation.


\bibitem{MW1}
Marinucci, D., Wigman, I.;  On the Excursion Sets of Spherical Gaussian Eigenfunctions,
available online http://arxiv.org/abs/1009.4367 .

\bibitem{MW2}
Marinucci, D., Wigman, I.; The defect variance of random spherical harmonics,
in preparation.


\bibitem{Neuheisel}
J. Neuheisel, {\em The asymptotic distribution of nodal sets on
  spheres}, Johns Hopkins Ph.D. thesis (2000).


\bibitem{NN}
Nonnenmacher, S.; Anatomy of quantum chaotic eigenstates, available online http://arxiv.org/abs/1005.5598 .


\bibitem{Nazarov Sodin}
Nazarov, Fedor; Sodin, Mikhail. On the number of nodal domains of random spherical harmonics. Amer. J. Math. 131 (2009), no. 5, 1337–1357.

\bibitem{Pleijel} A. Pleijel. Remarks on Courant's nodal line theorem, Comm. Pure Appl. Math. 9
(1956), 543--550.

\bibitem{RW}
 Z.~Rudnick and I.~Wigman
{\em On the volume of nodal sets for eigenfunctions of the Laplacian
on the torus}, Annales Henri Poincare, Vol. 9 (2008), No. 1, 109--130.

\bibitem{Smirnov}
Smirnov, Stanislav. Critical percolation in the plane: conformal invariance, Cardy's formula, scaling limits.
C. R. Acad. Sci. Paris Sér. I Math. 333 (2001), no. 3, 239–-244.

\bibitem{TW1}
Toth, John A.; Wigman, Igor {\em Counting open nodal lines of random waves on planar domains},
IMRN (2009).

\bibitem{Y1} Yau, S.T. Survey on partial differential equations in differential geometry.
{\it Seminar on Differential Geometry}, pp. 3--71, Ann. of Math. Stud., 102, Princeton Univ. Press, Princeton, N.J., 1982.


\bibitem{Y2} Yau, S.T. Open problems in geometry.
{\it  Differential geometry: partial differential equations on
manifolds} (Los Angeles, CA, 1990), 1--28, Proc. Sympos. Pure
Math., 54, Part 1, Amer. Math. Soc., Providence, RI, 1993.


\bibitem{Wig1}
Wigman, I. On the distribution of the nodal sets of random spherical harmonics. J. Math. Phys. 50 (2009), no. 1, 013521, 44 pp.

\bibitem{Wig2} Wigman, Igor Fluctuations of the nodal length of random spherical harmonics. Comm. Math. Phys. 298
(2010), no. 3, 787–831,

\bibitem{Wig3}
Wigman, I. {\em Volume fluctuations of the nodal sets of random Gaussian subordinated spherical harmonics},
unpublished.


\bibitem{Z} Zelditch, S. {\em Real and complex zeros of Riemannian random waves}.
To appear in the Proceedings of the Conference, "Spectral Analysis in Geometry and
Number Theory on the occasion of Toshikazu Sunada's 60th birthday", to appear in the Contemp. Math. Series,
available online http://arxiv.org/abs/0803.4334

\end{thebibliography}
\end{document}